\documentclass[iop,apj]{emulateapj}
\usepackage{amsmath,natbib,graphics,graphicx,apjfonts}

\input{buote_new.defs}
\newcommand\mlhband{{M_{\star}/L_H}}
\newcommand\src{\hbox{{Mrk~1216}}}

\newcommand\srctwo{\hbox{{PGC~032873}}}

\begin{document}

\title{The Luminous X-Ray Halos of Two Compact Elliptical Galaxies}
\author{David A.\ Buote and Aaron J.\ Barth}
\affil{Department of Physics and Astronomy, University of California
  at Irvine, 4129 Frederick Reines Hall, Irvine, CA 92697-4575; \\ buote@uci.edu}


\slugcomment{Accepted for Publication in The Astrophysical Journal}

\begin{abstract}
  There is mounting evidence that compact elliptical galaxies (CEGs)
  are ``massive relic galaxies'' that are local analogs of the
  high-redshift ``red nuggets'' thought to represent progenitors of
  today's early-type galaxies (ETGs).  We report the discovery of
  extended X-ray emission from a hot interstellar / intragroup medium
  in two CEGs, \src\ and \srctwo, using shallow archival \chandra\
  observations.  We find that \srctwo\ has an average gas temperature
  $k_BT=0.67\pm 0.06$~keV within a radius of 15~kpc of the galaxy,
  and a luminosity
  $L_{\rm x} = (1.8\pm 0.2)\times 10^{41}$~erg~s$^{-1}$ within a
  projected radius of 100~kpc, the latter of which is estimated by
  extrapolating the fitted $\beta$ model to the surface
  brightness. For \src, which is closer and more luminous
  [$L_{\rm x}(\rm <100~kpc) = (12.1\pm 1.9)\times
  10^{41}$~erg~s$^{-1}$], the data are of sufficient quality to
  perform a spatially resolved spectral analysis in seven circular
  annuli out to a radius of 73~kpc. Using an entropy-based hydrostatic
  equilibrium procedure, we obtain good constraints on the $H$-band
  stellar mass-to-light ratio, $M_{\rm stars}/L_H=1.33\pm 0.21$~solar,
  in good agreement with that obtained from stellar dynamical studies,
  which supports the hydrostatic equilibrium approximation for this
  galaxy. We obtain a mass-weighted density slope $2.22\pm 0.08$
  within $R_e$ consistent with other CEGs and normal local ETGs, while
  we find the dark matter (DM) fraction within $R_e$
  $f_{\rm DM}=0.20\pm 0.07$ to be similar to local ETGs. We place a
  constraint on the SMBH mass,
  $M_{\rm BH} = (5\pm 4)\times 10^{9}\, M_{\odot}$, with a 90\% upper
  limit of $M_{\rm BH} = 1.4\times 10^{10}\, M_{\odot}$, consistent
  with a recent stellar dynamical measurement. If we assume the
  Navarro-Frenk-White (NFW) DM scale radius does not lie beyond the
  current extent of the data, we also obtain interesting constraints
  on the halo concentration ($c_{200}=17.5\pm 6.7$) and mass
  [$M_{200} = (9.6\pm 3.7)\times 10^{12}\, M_{\odot}$]. The measured
  $c_{200}$ exceeds the mean \lcdm\ value ($\approx 7$), consistent
  with a system that formed earlier than the general halo
  population. We suggest that these galaxies, which reside in
  group-scale halos, should be classified as fossil groups.
\end{abstract}

\section{Introduction}
\label{intro}

Over the past decade it has become increasingly clear that most
early-type galaxies (ETGs) form and evolve via a two-stage
process~\citep[e.g.,][]{oser10a}. Initially, the galaxy collapses and
rapidly evolves through strong dissipation and wet mergers assembling
most of its stellar mass and becoming a ``red nugget'' by
$z\approx 2$. Compared to the present-day ETG population, red nuggets
are much more compact and have disky isophotes consistent with being
fast rotators. The second phase of ETG evolution is a gradual build-up
of its stellar envelope around the core red nugget through mostly dry
mergers and passive evolution of its stellar population. This is
reflected in the well-established size-mass evolution of
ETGs~\citep[e.g.,][]{dokkum08a} and through multi-component
decompositions of nearby ETGs~\citep{huan13b}. Unfortunately, because
red nuggets exist at high redshift it has not been possible yet to
measure radial mass profiles in detail to probe more effectively the
first phase.

A new way to approach studying red nuggets in more detail is through
local analogs. There is mounting evidence that compact elliptical
galaxies~(CEGs; e.g., \citealt{remco15a}) are indeed largely untouched,
passively evolved descendants of the high-redshift red nuggets. The
CEGs possess many of the same basic properties (e.g., small size,
large stellar mass, etc.), but only a relatively small number of CEGs
have been studied in detail. In the largest and most comprehensive
study to date, \citet[][hereafter Y17]{yild17a} present stellar
dynamical models of IFU kinematic data along with stellar populations
studies of 16 CEGs. Three of these have also been studied recently by
\citet{ferr17a} in detail confirming their identification as ``massive
relic galaxies'' (MRGs). \citet{yild17a} find that within $1R_e$ both
the total mass slope and the mean stellar mass fraction are higher
than present-day ETGs. They argue that both of these properties are
consistent with dissipative formation for the red nuggets.  Y17 also
argue their analysis of the CEGs disfavors adiabatic contraction of
their DM halos, which would represent an important constraint on the
ubiquity of that evolutionary
process~\citep[e.g.,][]{blum86a,gned04a,dutt15a}.

Finally, CEGs/MRGs have generated significant interest since several
studies suggest that they possess super-massive black holes (SMBHs)
that are positive outliers (i.e., \"{u}bermassive) at the high-mass
end $(>10^{9}\, M_{\odot})$ of the BH-mass scaling
relations~\citep[e.g.,][]{ferr15a,remco15a,wals15a,yild16a,wals17a},
similar to those BHs found in some
BCGs~\citep[e.g.][]{mcco12a,rusl13a,thom16a}.

What has been learned so far for the mass profiles has been
achieved only through stellar dynamics. These systems are too
nearby for studies with gravitational lensing. However, they are
potentially ideal sites for hydrostatic equilibrium (HE) studies of
their hot gas, given that they are believed to be largely untouched,
passively evolved descendants of the high-redshift universe. HE allows
the gravitating mass to be derived directly from the temperature and
density profiles of the ISM, from the center to the halo outskirts
using a single dynamical tracer.

In this paper we describe a search for promising CEG/MRG targets to
apply the HE approach to study the mass profiles to complement and
augment what has and is currently being learned from stellar
dynamics. In \S \ref{sample} we identify the CEG sample in which we
searched for targets with extended X-ray emission suitable for HE
analysis from which we identify two promising galaxies,
\src\footnote{Shortly before this paper was submitted for publication, a
  paper describing an analysis of gas heating and cooling using the
  \chandra\ data of \src\ was submitted to MNRAS and posted to
  arXiv.org by \citet{wern17a}.} and \srctwo. We describe the
\chandra\ X-ray observations and data preparation in \S \ref{obs}.  We
define the models used for spectral analysis in \S \ref{specmod}. In
\S \ref{pgc} we present results for \srctwo. For \src\ we describe the
models and results in \S \ref{mrk}. We present an analysis of the
image properties in \S \ref{image}, and the results of the spectral
analysis in \S \ref{spec}. The HE models are presented in \S \ref{he}
and the results in \S \ref{results}. We present our conclusions in \S
\ref{conc}.

\section{CEG Sample}
\label{sample}

\begin{table*}[t] \footnotesize
\begin{center}
\caption{Target Properties}
\label{tab.prop}
\begin{tabular}{lccccccccc}   \hline\hline\\[-7pt]
& & Distance & Scale & $N_{\rm H}$ & $L_{\rm IR}$ & $R_e$ & $\sigma_e$ & $L_{\rm x}$ & $\ktemp$ \\
Name & Redshift & (Mpc) & (kpc/arcsec) & ($10^{20}$~cm$^{-2}$) & $(10^{11}\, L_{\odot})$  & (kpc) & (km/s) & ($10^{41}$~ergs~s$^{-1}$) & (keV)\\
\hline \\[-7pt]
\src\ &  0.021328 & 97.0 & 0.45 & 4.0 & 1.14 & 2.3 & 308 & $12.1 \pm 1.9$ & $0.76\pm 0.02$\\
\srctwo\ & 0.024921 & 108.8 & 0.50 & 1.2 & 1.21 & 1.7 & 304 & $1.8 \pm 0.2$ & $0.67\pm 0.06$\\
\hline \\
\end{tabular}
\tablecomments{The redshift is taken from
  NED\footnote{http://ned.ipac.caltech.edu}. We compute the distance
  in our assumed cosmology using the redshift (also taken from NED)
  corrected to the reference frame defined by the 3K background.  We
  calculate the Galactic column density using the HEASARC {\sc w3nh}
  tool based on the data of~\citet{kalb05a}.  The total IR
  luminosities and circularized effective radii $(R_e$) are taken from
  Y17 for \src\ (H-band) and from 2MASS for \srctwo\ (K-band). The
  stellar velocity dispersions within $R_e$ are taken from Y17 for
  both systems.  $L_{\rm x}$ is the bolometric (0.1-50.0~keV)
  luminosity computed using the best-fitting hydrostatic model for each
  galaxy within an extrapolated projected radius 100~kpc (\S
  \ref{results}). The temperatures are average values computed within 73~kpc
  for \src\ (\S \ref{spec}) and 14.1~kpc for \srctwo\ (\S \ref{pgc}).}
\end{center}
\end{table*}

We searched for promising CEG targets for HE study of their hot ISM
using the new catalog of Y17 which presents stellar
dynamical studies of 16 nearby CEG. We summarize the results of our
initial survey of X-ray data archives as follows. 

\begin{itemize}

\item No \chandra\ or \xmm\ data -- NGC~384, NGC~472, NGC~2767, PGC~70520

\item Nuclear point sources with little extended emission -- UGC~2698
  and  UGC~3816 

\item ULX~\citep{walt11a} with little extended emission -- NGC~3990

\item Negligible / Insufficient diffuse emission --  PGC~011179,
  PGC~12562, NGC~1282

\item Perseus cluster galaxies for which determining the extended nature of emission
  is problematical due to the source being far off-axis, on a chip edge, and/or swamped
  by intracluster medium -- NGC~1270, NGC~1271, NGC~1277, NGC~1281
 
\item Isolated galaxies with luminous, extended X-ray emission --  \src\
  and \srctwo

\end{itemize}

Unfortunately, most of the CEGs in the Y17 catalog are not promising
for study presently for a variety of reasons. Some lack any \chandra\ or
\xmm\ data to search for extended emission. Most of the targets do not
show clear evidence for substantial extended emission with the
existing data. However, two targets are very promising for HE study of
extended X-ray emission: (1) \src\ and (2) \srctwo. Both of these
objects are recently very well-studied in optical/IR confirming their
status as MRGs~\citep{ferr17a,yild17a} and possibly with over-massive
SMBHs~\citep[e.g.,][]{ferr15a,wals17a}. We list their basic properties
in Table \ref{tab.prop} and below analyze in detail their extended
X-ray emission.

\section{Observations and Data Preparation}
\label{obs}

\begin{table}[t] \footnotesize
\begin{center}
\caption{Observations}
\label{tab.obs}
\begin{tabular}{ccccc}   \hline\hline\\[-7pt]
& & & & Exposure\\
Galaxy & Obs.\ ID & Obs.\ Date & Instrument & (ks)\\
\hline \\[-7pt]
\src\ & 17061 & 2015  Jun.\ 12 & ACIS-S & 12.8\\
\srctwo\ & 17063 & 2015 Mar.\ 2 & ACIS-S & 22.7\\
\hline \\
\end{tabular}
\tablecomments{The exposure times refer to those obtained after
  filtering the light curves, which for each galaxy
  resulted in $<1$~ks of excluded time.}
\end{center}
\end{table}

In Table~\ref{tab.obs} we list details of the \chandra\ observations
of \src\ and \srctwo.  Unless stated otherwise, the data were prepared
as described in \citet[][hererafter B17]{buot17a}, and we refer the
reader to that paper for details.  We used the \ciao\ (v4.9) and
\heasoft\ (v6.18) software suites along with version 4.7.5.1 of the
\chandra\ calibration database to prepare the data for imaging and
spatially resolved spectral analysis. For the imaging analysis, we
extracted images from the cleaned events lists with energies
$0.5-7.0$~keV and employed 1.7~keV monochromatic exposure maps.

For the spectral analysis, we required a minimum 200 source counts for
each concentric circular annulus. In the case of \srctwo, this
resulted in only a single aperture ($R=30\arcsec=15.1$~kpc). We also
included a larger annulus ($R=30\arcsec-150\arcsec$) to help constrain
the background.  For \src\ our criterion resulted in seven annuli (see
Table~\ref{tab.gas}) extending out to 73~kpc ($2\farcm 7$) plus a
larger annulus (2\farcm 7-4\farcm 1) for aiding background
constraints.

We also examined the likelihood that enhanced Solar Wind Charge
Exchange (SWCX) emission significantly impacted the observations. We
obtained the solar proton flux during each \chandra\ observation using
the Level~3 data from
SWEPAM/SWICS\footnote{http://www.srl.caltech.edu/ACE/ASC/level2/sweswi\_l3desc.html}.
Both observations have solar proton flux below
$\approx 2\times 10^8$~cm$^{-2}$~s$^{-1}$ indicating significant
proton flare contamination is not expected~\citep{fuji07a}.

\section{Spectral Models}
\label{specmod}

We performed frequentist fits of the plasma and background emission
models to the spectra using \xspec\ v12.9.0s~\citep{xspec}. We chose
to minimize the C-statistic~\citep{cstat} since it is largely unbiased
compared to $\chi^2$~\citep[e.g.][]{hump09a}. We modeled the hot ISM
with the \vapec\ plasma emission model and the cosmic and particle
backgrounds with a combination of power-laws and gaussians.  For both
galaxies the unresolved LMXB contribution is unimportant, and we
modeled it as a 7.3~keV thermal bremsstrahlung component~\citep[e.g.,][]{mats97a,irwi03a} with
normalization fixed by the $L_x - L_K$ scaling relation for unresolved
discrete sources of \citet{hump08b} using the $K$-band luminosity
listed in Table~\ref{tab.prop}.  The hot ISM components are allowed to
vary between the annuli while the background components are tied. For
each galaxy, the background models are also fitted in the extra large
annulus (\S \ref{obs}), which provides the key constraints on the
background. We refer the reader to \S 3 of B17 for details of the
models and fitting procedure.

For \src\ we found the soft Cosmic X-ray Background (CXB) components
fitted to negligible fluxes with large errors. Consequently, for that
galaxy we fixed the soft CXB normalizations to those obtained from
fitting {\sl ROSAT} data using the HEASARC X-ray Background
Tool\footnote{https://heasarc.gsfc.nasa.gov/cgi-bin/Tools/xraybg/xraybg.pl}.

\section{PGC~032873}
\label{pgc}

\begin{figure*}
\parbox{0.49\textwidth}{ 
\centerline{\includegraphics[scale=0.35,angle=0]{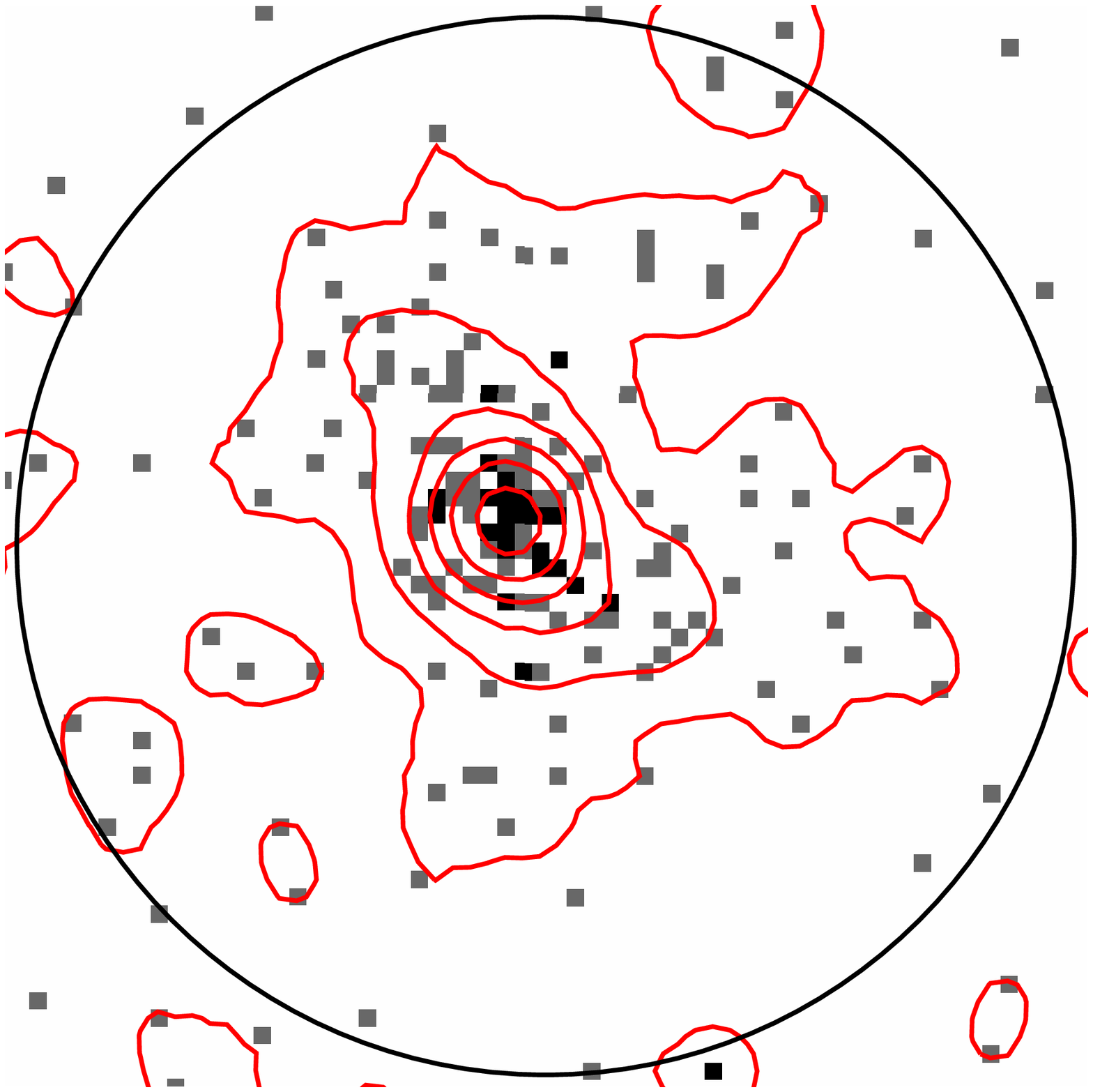}}}
\parbox{0.49\textwidth}{
\centerline{\includegraphics[scale=0.35,angle=-90]{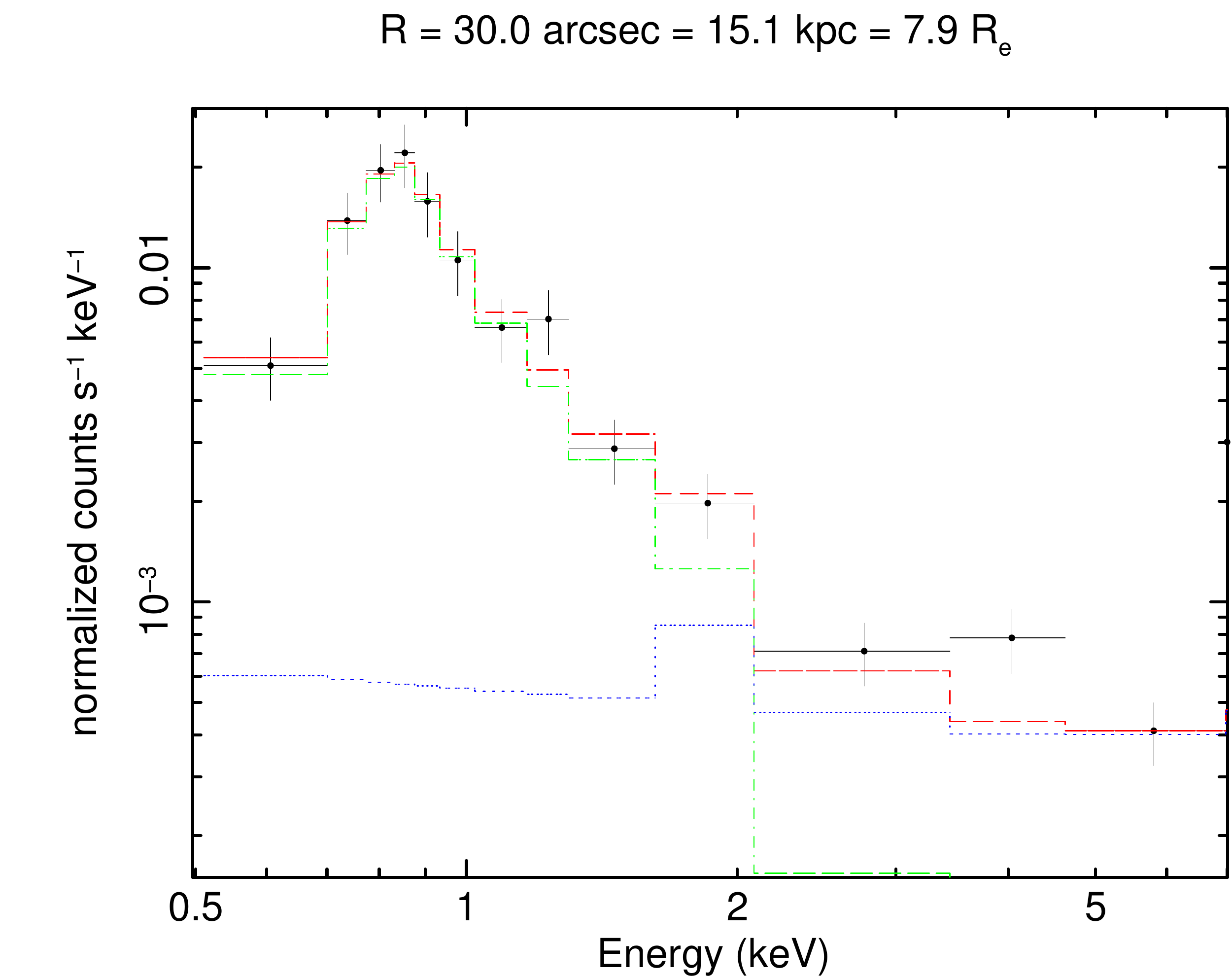}}}
\caption{\label{fig.pgc.image} Image and spectrum of \srctwo.  ({\sl
    Left Panel} ) \chandra\ image (0.5-7.0~keV, $0\farcs 492$ pixels)
  with contours overlaid with square-root spacing.  Also shown is a
  circle of radius $15\arcsec=7.5$~kpc for scale. Note this is the
  raw image used only for display purposes; i.e., no exposure
  correction or background-subtraction has been applied. ({\sl Right
    Panel}) \chandra\ spectrum extracted within a radius of
  $30\arcsec$. The best-fitting spectral model is shown: hot gas +
  unresolved discrete sources (LMXBs) + CXB (green), particle
  background (blue), total model (red).}
\end{figure*}

\begin{figure*}[t]
\begin{center}
\includegraphics[scale=0.42,angle=-90]{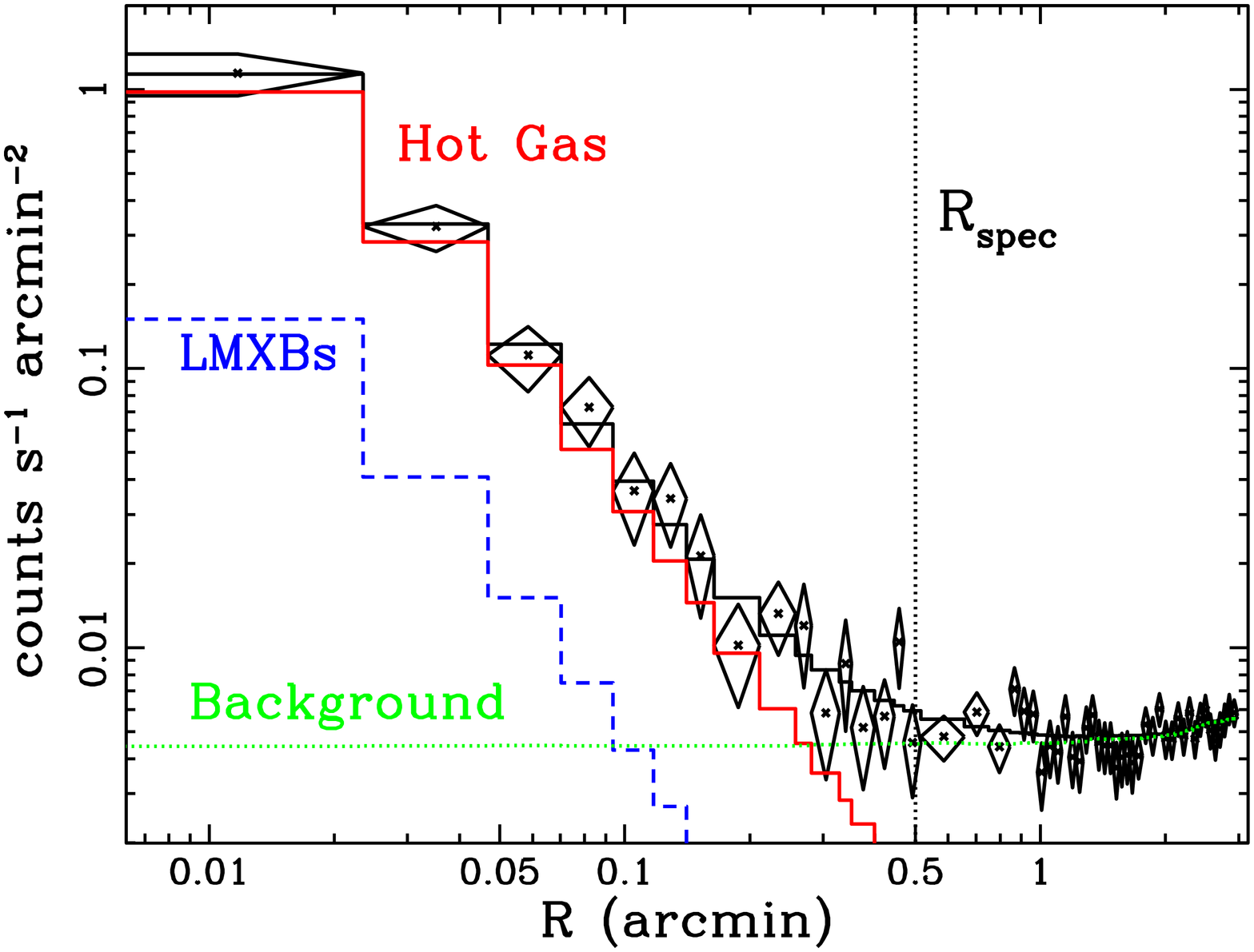}
\end{center}
\caption{\footnotesize 0.5-7.0~keV radial surface brightness profile
  and best-fitting model for \srctwo\ (see \S \ref{pgc}).  The vertical
  dotted line indicates the spectral extraction aperture at radius
  $R_{\rm spec}=30\arcsec=15.1$~kpc. }
\label{fig.pgc.sb}
\end{figure*}

In Figure~\ref{fig.pgc.image} we display the image overlaid with
contours for the central region ($\approx 15\arcsec$) of
\srctwo. Although the number of source counts is small (only $\approx
170$ within the displayed circle), extended emission centered on the
stellar light is clearly observed.  The morphology of the X-ray
isophotes is broadly consistent with that of the $H$-band image
reported by Y17; i.e., isophotal ellipticity 0.53 and position angle
$42^{\circ}$. (In Figure ~\ref{fig.pgc.image} North is up and East is to
the left.)

We also plot in Figure~\ref{fig.pgc.image} the spectrum and
best-fitting model in a circular aperture with radius
$R=30\arcsec = 15.1$~kpc containing $\approx 200$ source counts. The
temperature of the hot ISM within the aperture is well constrained,
$k_BT = 0.67\pm 0.06$~keV with a sub-solar (though less certain)
metallicity, $Z=0.42\pm 0.23Z_{\odot}$.  The best-fitting aperture
luminosity is $L_{\rm x}=6.8\times 10^{40}$~erg~s$^{-1}$
(0.5-7.0~keV). These properties are indicative of a typical X-ray
luminous massive elliptical galaxy~\citep[e.g.,][]{hump06b}; e.g.,
using the mass-temperature scaling relation of galaxy groups
from~\citet{lovi15a} gives $M_{500}=2\times 10^{13}\, M_{\odot}$.

Since the data are insufficient for spectral analysis in multiple
apertures, we also estimate the radial mass profile using the surface
brightness profile and approximating the gas as isothermal. In
Figure~\ref{fig.pgc.sb} we show the 0.5-7.0~keV radial surface
brightness profile out to a radius of $3\arcmin$. We fitted a model to
the surface brightness consisting of (1) an isothermal $\beta$
model~\citep{beta} for the hot gas; (2) a de Vaucouleurs model
following the $K$ band light (Table~\ref{tab.prop}) normalized as
described in \S \ref{specmod} to represent the unresolved LMXB
component; and (3) two constant background components, one of which
represents the particle background and is not corrected by the
exposure map.

The composite model is a good fit to the surface brightness profile
and yields for the $\beta$ model a core radius
$0\farcs 6\pm 0\farcs 3$, slope parameter
$\beta=0.50^{+0.03}_{-0.02}$, and best-fit central density,
$\rho_{\rm gas,0} = 9.5\times 10^{-25}$~g~cm$^{-3}$. Assuming
HE and using these best-fitting parameters along
with the temperature and metallicity obtained for the spectrum quoted
above, the best-fit $\beta$ model profile yields a total mass,
$M_{500}=1.3\times 10^{13}\, M_{\odot}$, very consistent with the
value obtained above from a scaling relation when considering the
large statistical uncertainties.

\section{Mrk~1216}
\label{mrk}

\subsection{Image Properties}
\label{image}


\begin{figure*}
\parbox{0.49\textwidth}{ 
\centerline{\includegraphics[scale=0.35,angle=0]{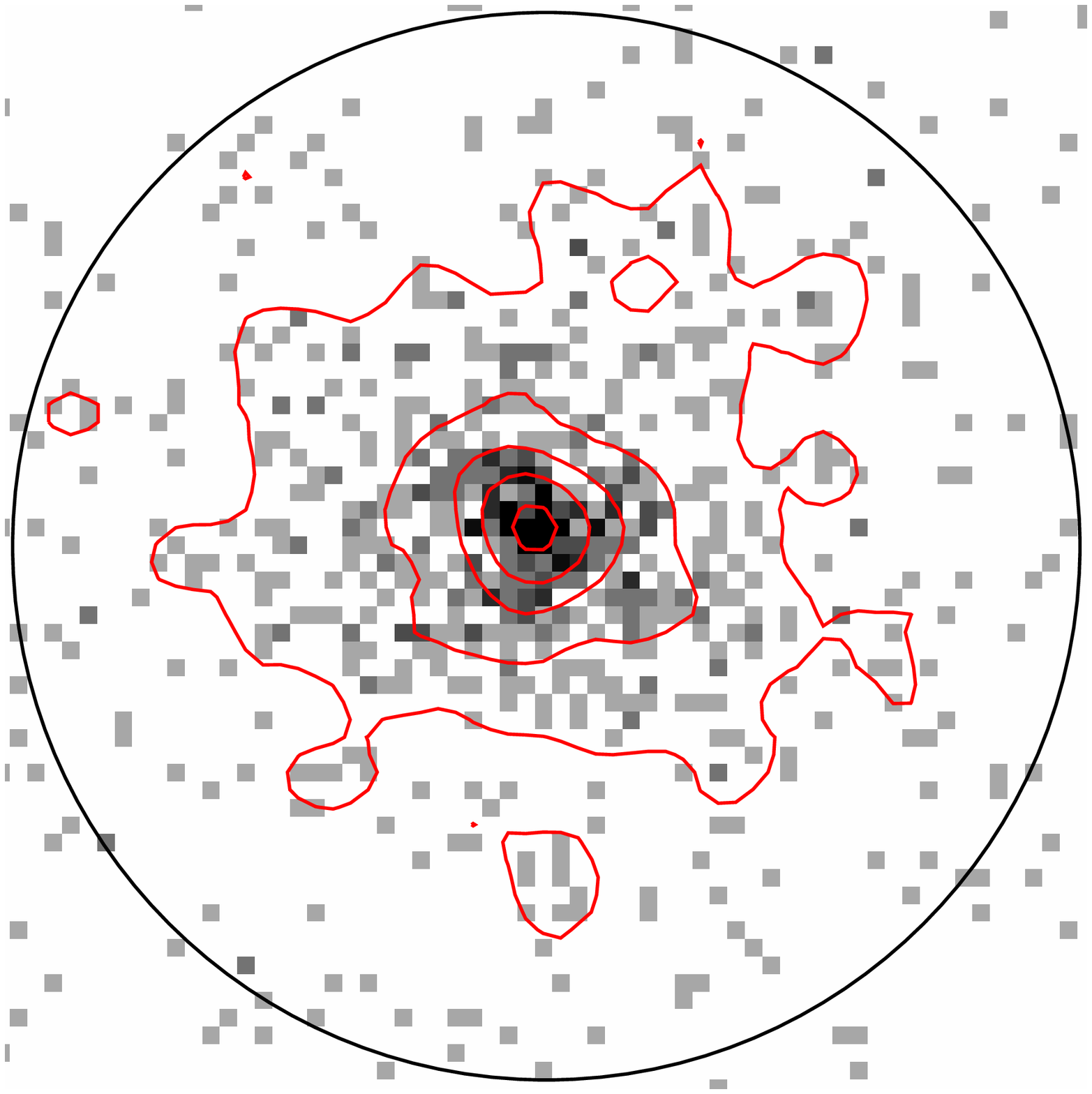}}}
\parbox{0.49\textwidth}{ 
\centerline{\includegraphics[scale=0.325,angle=0]{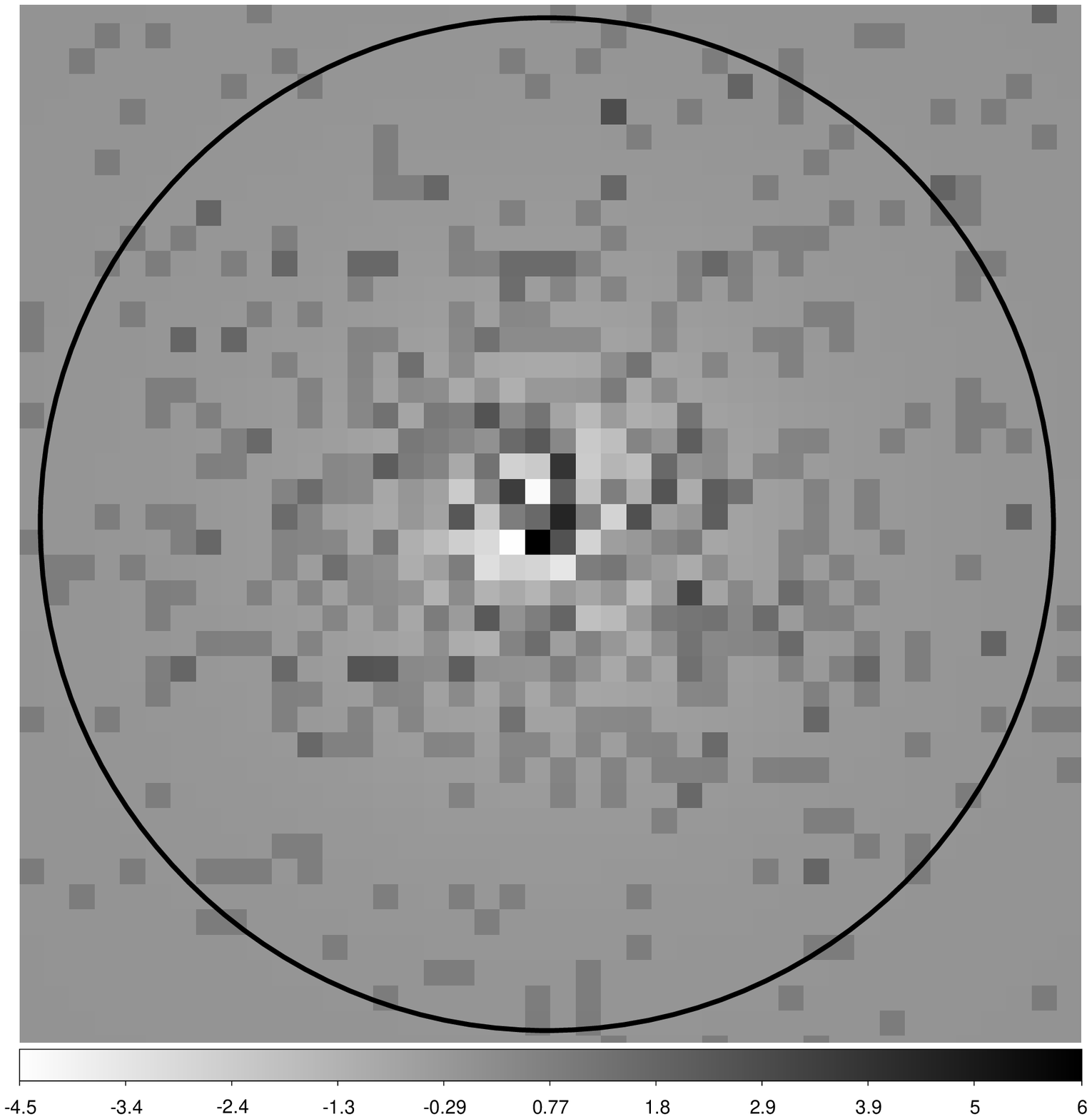}}}
\vskip 1.5cm
\caption{\label{fig.image} 0.5-7.0~keV \chandra\ image ($0\farcs 492$
  pixels) and residuals for \src.  ({\sl Left Panel} ) Image with
  contours overlaid with square-root spacing.  Also shown is a circle
  of radius $15\arcsec=6.7$~kpc for scale. Note this is the raw image
  used only for display purposes; i.e., no exposure correction or
  background-subtraction has been applied.  ({\sl Right Panel})
  Residuals obtained after subtracting the best-fitting circular
  $\beta$-model fitted to the exposure-corrected image. (The displayed
  circle has a radius of $10\arcsec$.)  The spatial fluctuations near
  the galaxy's center are not statistically significant. }
\end{figure*}

We show the 0.5-7.0~keV \chandra\ image of the central
$\sim 15\arcsec$ of \src\ in Figure~\ref{fig.image} with contours
overlaid. Like \srctwo, the X-ray image of \src\ shows its emission is
clearly extended and is centered on the peak of the stellar light.
The displayed region contains $\approx 750$ source counts allowing for
a more detailed analysis than was possible for \srctwo. After
identifying point sources with the \ciao\ {\sc wavdetect} tool we
replaced them with local background using the \ciao\ {\sc dmfilth}
tool. We then computed the ellipticity and position angle of the
surface brightness brightness as a function of semi-major axis using
an iterative moment technique equivalent to diagonalizing the moment
of inertia tensor (\citealt{cm}; see \citealt{buot94} for application
to X-ray images of elliptical galaxies).

Within the displayed region the X-ray position angle (PA, measured N
through E) is consistent with following the $H$-band stellar light
reported by Y17 ($70.5^{\circ}$), while the ellipticity is smaller
than the $H$-band value $(\epsilon=0.42)$; e.g., for semi-major axis
$10\arcsec$ we obtain $\epsilon = 0.24\pm 0.07$ and
$\rm PA = 71^{\circ}\pm 11^{\circ}$. The rounder X-ray isophotes
compared to the stars suggests the X-ray emission follows the
gravitational potential obeying approximate HE. This is further
supported by the lack of any centroid variation.

To search for image irregularities in more detail, we used the \ciao\
package {\sc sherpa} to fit a circular $\beta$ model to the image and
subtract the best-fitting model yielding the residual image shown in
the right panel of Figure~\ref{fig.image}.  (Note that for this
calculation the image was first corrected with a 1.5~keV monochromatic
exposure map. However, we found the results were affected negligibly
whether or not the exposure correction was applied.)  We do not find any
statistically significant features in the residual image. Hence, with
the present data, the X-ray emission of \src\ appears to be very
regular and consistent with a relaxed system.

\subsection{Spectral Analysis}
\label{spec}

\begin{figure*}
\parbox{0.49\textwidth}{
\centerline{\includegraphics[scale=0.35,angle=-90]{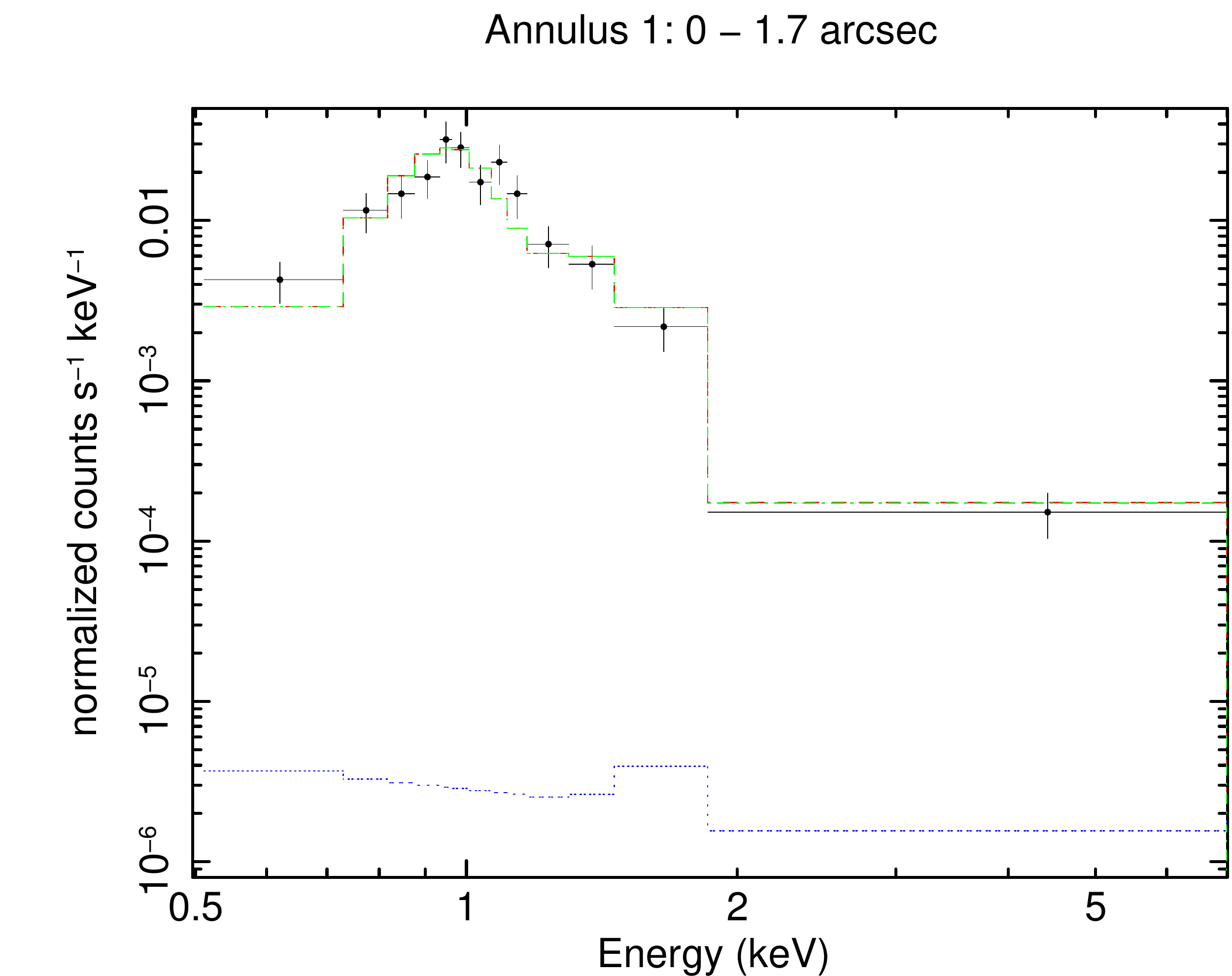}}}
\parbox{0.49\textwidth}{
\centerline{\includegraphics[scale=0.35,angle=-90]{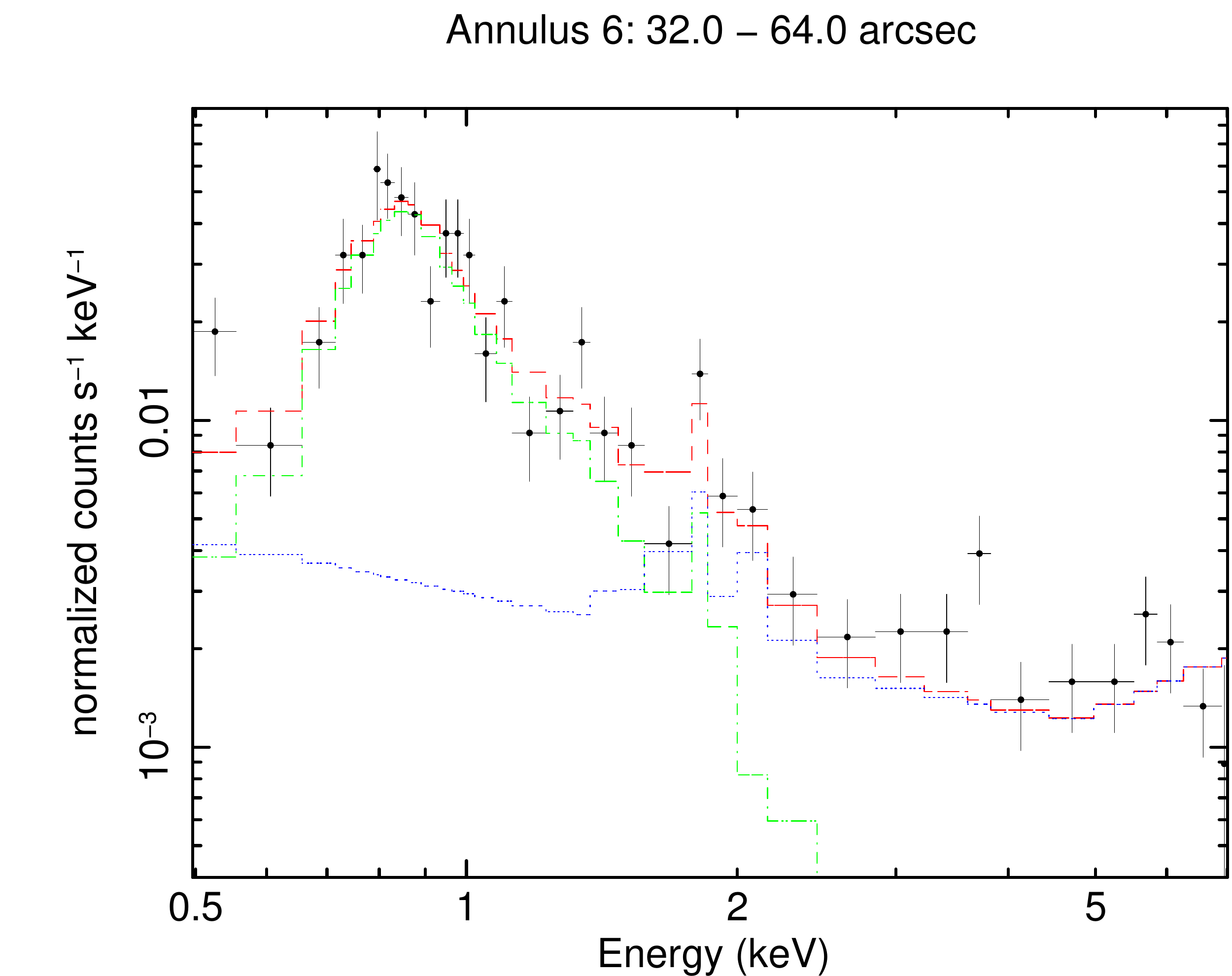}}}
\caption{\label{fig.spec} 
 Example \chandra\ spectra for \src\ in the
  0.5-7.0~keV band without any background subtraction. Also plotted
  are the best-fitting models (red dashed) broken down into the separate contributions
  from the following:  (1) hot gas and unresolved LMXBs from \src\
  along with the CXB (green dot-dash), and (2) particle background
  (blue dotted).}
\end{figure*}

\begin{figure*}
\parbox{0.49\textwidth}{
\centerline{\includegraphics[scale=0.43,angle=0]{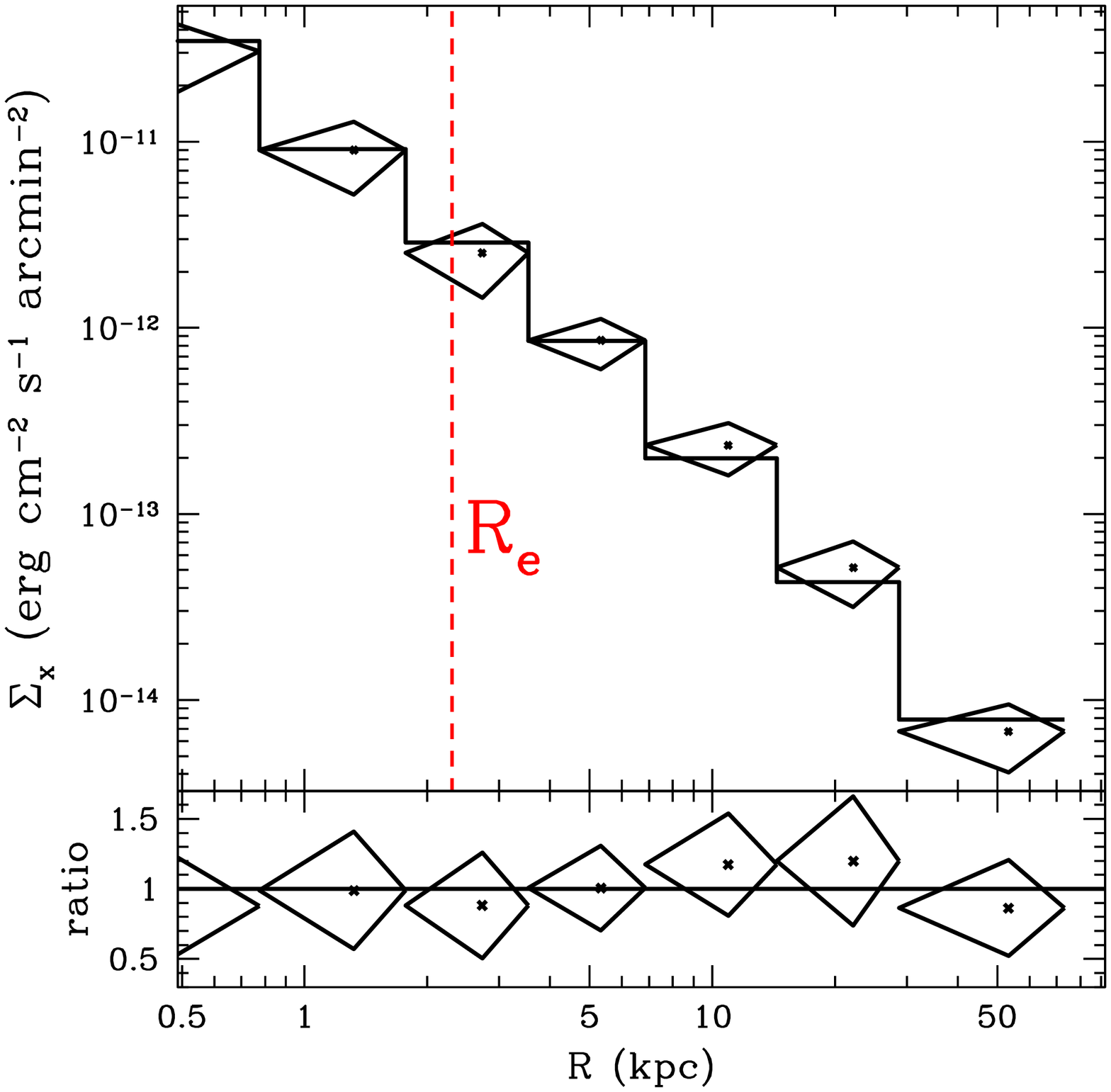}}}
\parbox{0.49\textwidth}{
\centerline{\includegraphics[scale=0.43,angle=0]{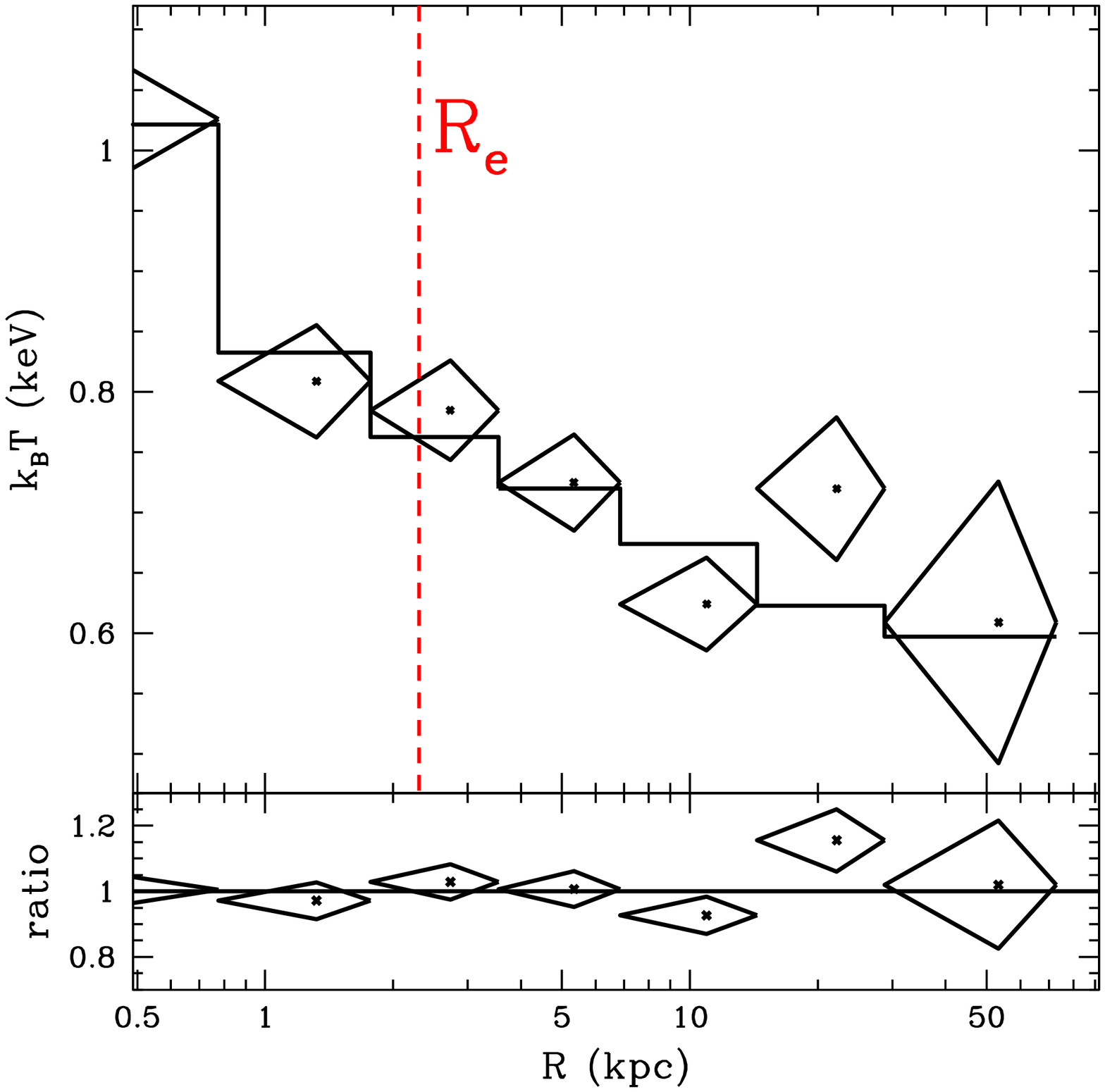}}}
\caption{\label{fig.best} \chandra\ data, $1\sigma$ errors, and the
  best-fitting fiducial hydrostatic model in each circular annulus on
  the sky for \src.  ({\sl Left Panel}) Surface brightness
  (0.5-7.0~keV). See the notes to Table~\ref{tab.gas} regarding the
  error bars on $\xsurf$.  ({\sl Right Panel}) Projected
  emission-weighted temperature ($k_BT$). Also shown is the location
  of the stellar half-light radius ($R_e$). The bottom panels plot the
  data/model ratios.}
\end{figure*}

\begin{table*}[t] \footnotesize
\begin{center}
\caption{Hot Gas Properties for \src}
\label{tab.gas}
\begin{tabular}{cccccccc}   \hline\hline\\[-7pt]
 & $R_{\rm in}$ & $R_{\rm out}$ & $\Sigma_{\rm x}$ (0.5-7.0~keV) & $k_BT$ & $Z$\\
 Annulus & (kpc) & (kpc) & (ergs cm$^2$ s$^{-1}$ arcmin$^{-2}$) & (keV) & (solar)\\
\hline \\[-7pt]
  1 & 0.00 & 0.78 &   3.06e-11 $\pm$   1.21e-11 & $  1.026 \pm   0.041$ &  $   1.04 \pm    0.58$ \\
  2 & 0.78 & 1.77 &   9.00e-12 $\pm$   3.81e-12 & $  0.809 \pm   0.046$ &  tied \\
  3 & 1.77 & 3.54 &   2.52e-12 $\pm$   1.08e-12 & $  0.785 \pm   0.041$ &  tied \\
  4 & 3.54 & 6.86 &   8.55e-13 $\pm$   2.57e-13 & $  0.725 \pm   0.040$ &  $   0.65 \pm    0.34$ \\
  5 & 6.86 & 14.38 &   2.34e-13 $\pm$   7.28e-14 & $  0.624 \pm   0.038$ &  tied \\
  6 & 14.38 & 28.77 &   5.15e-14 $\pm$   1.98e-14 & $  0.720 \pm   0.059$ &  $   0.38 \pm    0.21$ \\
  7 & 28.77 & 73.03 &   6.78e-15 $\pm$   2.68e-15 & $  0.609 \pm   0.117$ &  tied \\
\hline \\
\end{tabular}
\tablecomments{1~kpc = 2.22\arcsec. Annuli where the metallicity is
  linked to the value in the previous annulus are indicated as
  ``tied.''  Note that the definition of $\xsurf$ is essentially the
  emission measure (i.e., \xspec\ {\sc norm} parameter, which is the
  parameter actually fitted to the spectral data) multiplied by the
  plasma emissivity divided by $\pi\theta^2$ (arcmin$^2$), where
  $\theta$ is the aperture radius in arcminutes. Rather than quote the results for
  {\sc norm} itself, we have used the best-fitting plasma emissivity for each
  annulus (i.e., the plasma emissivity evaluated using the
  best-fitting $\ktemp$ and metallicity $Z$) to convert {\sc norm} into a
  surface brightness unit. Consequently, the error bars quoted for
  $\xsurf$ are directly proportional to the error bars for {\sc
    norm}.}
\end{center}
\end{table*}

We obtain a good joint fit to the spectra in the seven annuli with a
minimum C-statistic of 318 for 289 degrees of freedom (dof). The
$\chi^2$ value for this fit is 312 yielding a null hypothesis
probability of 17\% for a formally acceptable fit.  The good quality
of the global fit is apparent in Figure~\ref{fig.spec} where we show
the best-fitting model over-plotted on the spectra in two of the
annuli.

In Table~\ref{tab.gas} we list the surface brightness ($\xsurf$),
temperature ($\ktemp$), and the metallicity ($Z$) for the hot gas
component in each annulus. The metallicity parameter is defined such
that all the metal abundances other than iron are tied to iron in
their solar ratios, where we use the solar abundance table of
\citet{aspl}.  The radial profiles of $\xsurf$ and $\ktemp$ are
plotted in Figure~\ref{fig.best}.

The temperature declines from a maximum value of $\approx 1$~keV in
the central radial bin to $\approx 0.6$~keV in the outermost aperture
very similar to the temperature profile of NGC~6482 (B17). The source
emission in the central bin where the temperature peaks is well
described by thermal plasma emission. We found no evidence for
spectrally hard non-thermal emission in the central bin when adding a
power-law component potentially associated with the weak radio source
detected in the NVSS~\citep{nvss}.  (Since in \S \ref{conc} we will
have use of the average temperature, when tying $\ktemp$ for all the
apertures we obtain, $\ktemp = 0.76\pm 0.02$~keV.)

The metallicity is consistent with a significant negative radial
gradient; i.e., $Z$ declines from $Z\approx 1\, Z_{\odot}$ at the
center to $Z\approx 0.4\, Z_{\odot}$ in the outer radial bin similar
to NGC~6482 (B17), NGC~5044~\citep{buot03b} and other massive
elliptical galaxies and small
groups~\citep[e.g.,][]{buot00c,hump06a,mern17a}.  While the data are
consistent with the metallicity gradient, it is not required. A fit of
similar quality is obtained for a constant
$Z=0.73^{+0.23}_{-0.16}\, Z_{\odot}$. The statistical errors on the
iron abundance (which dominates the metallicity) are sufficiently
large to render unimportant any Fe bias~\citep[e.g.,][]{buot00a} from
fitting a single temperature model to a multi-temperature spectrum
(such as that arising from the line-of-sight projection of a radial
temperature gradient). We therefore focus our analysis on
single-temperature models of the hot ISM in each annulus

We also found that allowing other elements (i.e., O, Ne, Mg, Si, S) to
vary separately from Fe resulted in little improvement in the fit and
poorly constrained non-Fe abundances. The strongest statement that we
can report regarding these is that the ratios $Z_{\rm Mg}/Z_{\rm Fe}$
and $Z_{\rm Si}/Z_{\rm Fe}$ are less than solar at 90\% confidence.

\subsection{Hydrostatic Equilibrium Models}
\label{he}

We adopt a bayesian entropy-based procedure to fit spherical HE models
of the hot ISM to the \chandra\ spectral data (\citealt{hump08a}; see
\citealt{buot12a} for a review of this and other HE approaches). The
biases associated with assuming spherical symmetry are small
generally~\citep[e.g.,][]{buot12c}, and they are negligible in our
present investigation relative to the large statistical errors on
the fitted parameters. We refer the reader to B17 for details of the
implementation of the method. Here we briefly summarize the fiducial
model components used for \src.

\begin{itemize}

\item{\bf Entropy } Power-law plus a constant,
  $S(r)=s_0 + s_1r^\alpha$, where $S \equiv k_{\rm B}Tn_e^{-2/3}$ is
  the entropy proxy expressed in units of keV~cm$^{2}$, $r$ is
  expressed in kpc, and $s_0$, $s_1$, and $\alpha$ are free
  parameters. We also require at some radius outside the data extent
  that the logarithmic entropy slope match the value $\approx 1.1$
  from cosmological simulations with only
  gravity~\citep[e.g.,][]{tozz01a,voit05a}.  We adopt a radius of
  100~kpc for this purpose.

\item{\bf Pressure } The pressure boundary condition for the solution
  of the HE equation is a free parameter. We designate this
  ``reference pressure'' $P_{\rm ref}$ to be located at a radius
  10~kpc.

\item{\bf Stellar Mass } Multi-gauss expansion (MGE)
  model of the {\sl HST} $H$-band light reported by Y17. This stellar
  light profile is converted to stellar mass via the stellar
  mass-to-light ratio ($M_{\rm stars}/L_{\rm H}$), which is a free
  parameter in our model. 

\item{\bf Dark Matter } NFW
  profile~\citep{nfw} with free parameters concentration and mass. 

\item{\bf SMBH } We fix $M_{\rm BH}=4.9\times 10^9\,
  M_{\odot}$ to the stellar dynamical value~\citep{wals17a} for the fiducial HE model and discuss
  results obtained for  other $M_{\rm BH}$ values in \S \ref{smbh}.

\end{itemize}

Hence, our fiducial HE model has three free parameters for entropy, one
for pressure, one for stellar mass, and two for the DM; i.e., a total
of seven free parameters. 

\subsection{Results}
\label{results}

\begin{figure*}
\parbox{0.49\textwidth}{
\centerline{\includegraphics[scale=0.43,angle=0]{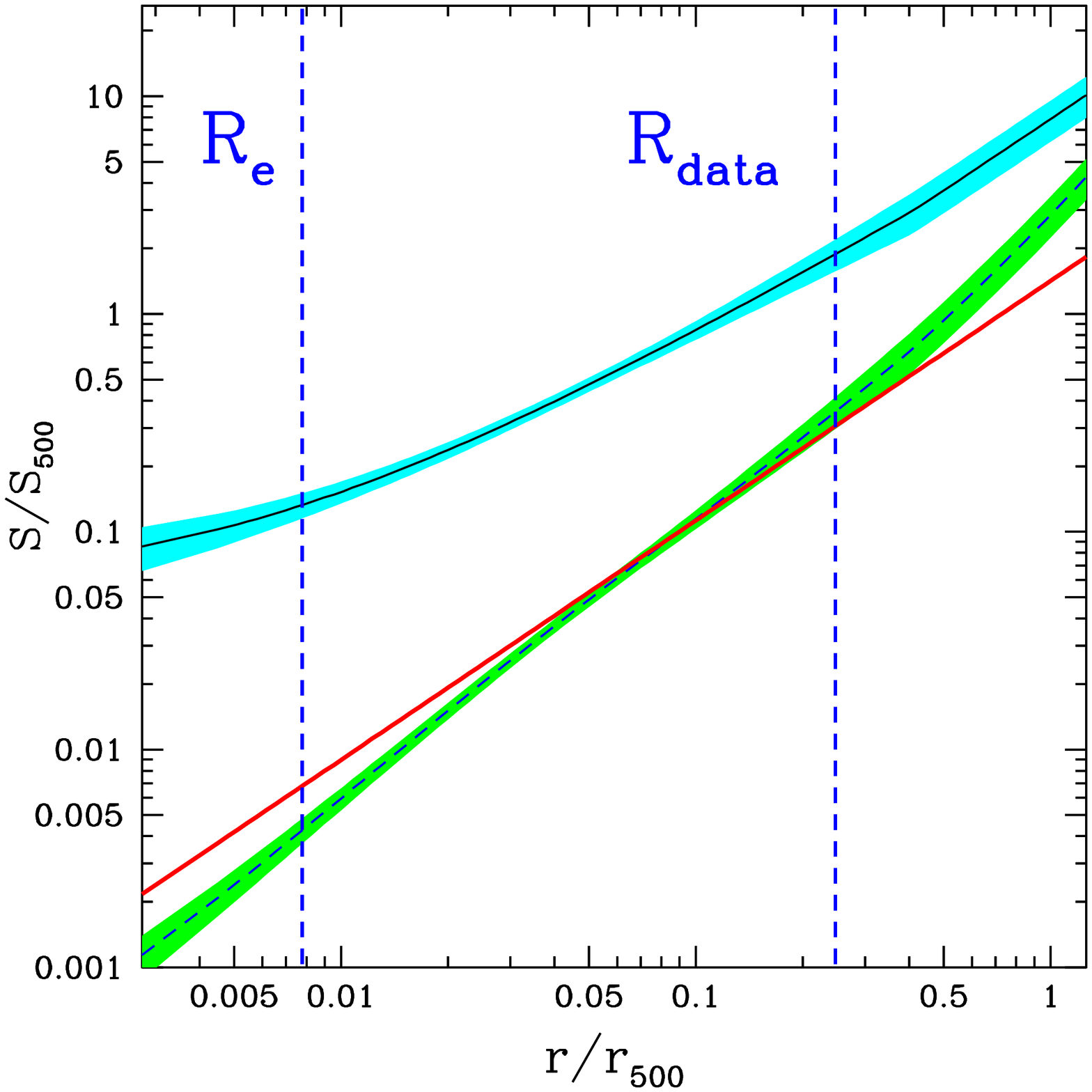}}}
\parbox{0.49\textwidth}{
\centerline{\includegraphics[scale=0.43,angle=0]{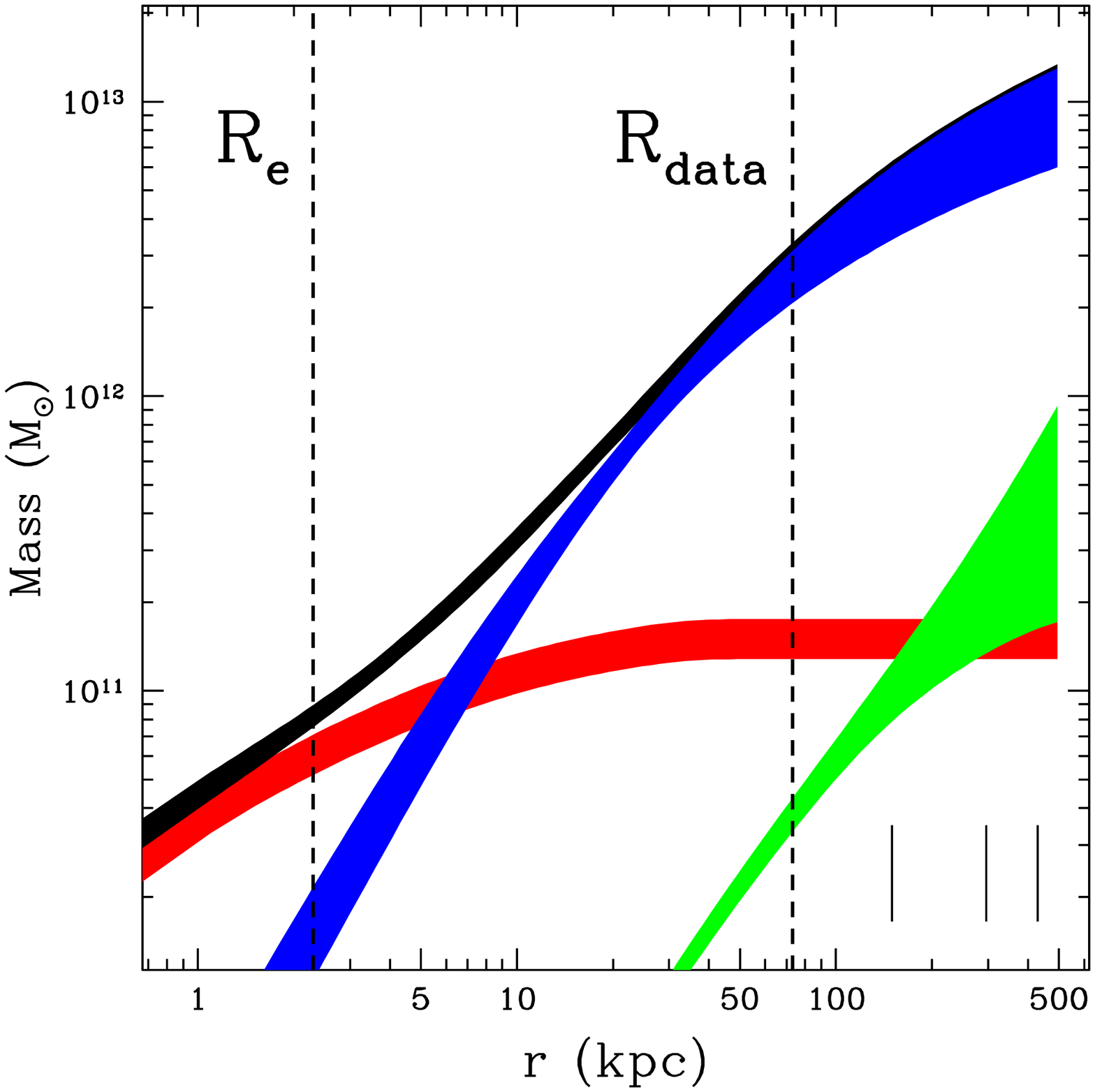}}}
\caption{\label{fig.he} Results for the HE modeling of \src.  The
  curved lines and associated shaded regions in both plots show the
  mean and standard deviation of the posterior as a function of radius
  for the quantity of interest; i.e., entropy or mass.  ({\sl Left
    Panel}) Radial profile of the entropy (black) and $1\sigma$ error
  region (cyan) for the fiducial hydrostatic model rescaled by
  $S_{500}=46.1$~keV~cm$^2$. The baseline $r^{1.1}$ profile obtained
  by cosmological simulations~\citep{voit05a} with only gravity is
  shown as a red line. The result of rescaling the entropy profile by
  $\propto f_{\rm gas}^{2/3}$ \citet{prat10a} is shown by the black
  dashed line (and green $1\sigma$ region). ({\sl Right Panel}) Radial
  profiles of the total mass (black) and individual mass components of
  the fiducial hydrostatic model: total NFW DM (blue), stars (red),
  hot gas (green).  The black vertical lines in the bottom right
  corner indicate the virial radii; i.e., from left to right:
  $r_{2500}$, $r_{500}$, and $r_{200}$. The vertical dashed lines
  indicate the location of the stellar half-light radius $(R_e)$ and
  the outer extent of the \chandra\ data analyzed $(R_{\rm data})$.}
\end{figure*}

\begin{table*}[t] \footnotesize
\begin{center}
\caption{Results for HE Mass Model of \src}
\label{tab.he}
\begin{tabular}{lc|ccc|c|cc|cc}  \hline\hline\\[-7pt]
 & $P_{\rm ref}$ & $s_0$ & $s_1$ & $\alpha$ & $M_{\star}/L_H$ & $c_{200}$ & $M_{200}$ & $f_{\rm gas, 200}$ & $f_{\rm b, 200}$ \\ 
& ($10^{-2}$ keV cm$^{-2}$) & (keV cm$^2$) & (keV cm$^2$) & & ($M_{\odot} L_{\odot}^{-1}$) &  & $(10^{12}\, M_{\odot})$\\ 
\hline \\[-7pt]
Best Fit & $0.98 \pm 0.10$ & $2.49 \pm 1.19$ & $2.07 \pm 0.71$ & $0.92 \pm 0.11$ & $1.33 \pm 0.21$ & $17.5 \pm  6.7$ & $ 9.6 \pm  3.7$  & $0.051 \pm 0.039$ & $0.070 \pm 0.045$ \\
(Max Like) & $(0.95)$ & $(1.31)$ & $(2.40)$ & $(0.86)$ & $(1.36)$ & $(25.9)$ & $( 5.1)$ & $(0.106)$ & $(0.137)$\\
\hline \\
\end{tabular}
\tablecomments{Best values and error estimates (see \S \ref{overview})
  for the fundamental free parameters of the fiducial HE
  model. $P_{\rm ref}$ refers to the total gas pressure evaluated at
  the reference radius $r=10$~kpc and serves as the boundary condition
  for the hydrostatic model. The parameters $s_0$, $s_1$, and
  $\alpha_1$ specify the power-law plus constant entropy profile. The
  fundamental mass parameters are the $H$-band stellar mass-to-light
  ratio $(M_{\star}/L_H)$, and the concentration and enclosed total mass
  (stars+gas+DM) computed within $r_{200}$. Note the gas and baryon
  fractions are parameters derived from the mass model.}
\end{center}
\end{table*}

\subsubsection{Overview}
\label{overview}

We use a bayesian nested sampling procedure based on the MultiNest
code v2.18~\citep{multinest} to fit the HE model to the \chandra\ data
(see B17 for details).  For the free parameters we use flat priors
except for $P_{\rm ref}$ and ($M_{\rm DM}$) for which we adopt flat
priors on their logarithms. The ranges of the priors were chosen to be
large enough so that the best-fitting values were far from the
boundaries as judged by the standard deviation of the parameter. The
one exception to this is the NFW scale radius $(r_s)$ for which the
upper limit is poorly constrained. Consequently, we set the maximum
value of the prior for $r_s$ to 50~kpc representing essentially the
average radius of the outer \chandra\ bin.

We  quote two ``best'' values for each free parameter: ``Best Fit'',
the mean parameter value of the posterior, and ``Max Like'', the
parameter value that maximizes the likelihood. Errors quoted are the
standard deviation ($1\sigma$) of the posterior unless stated
otherwise. In Figure~\ref{fig.best} we show the best-fitting fiducial
model to the $\xsurf$ and $\ktemp$ profiles and the fractional
residuals. The fit is excellent as judged by the small fractional
residuals. We have also performed a standard frequentist $\chi^2$
analysis to provide another means of judging the goodness-of-fit. This
fit yields parameters extremely similar to the ``Max Like'' parameters
of the bayesian fit and $\chi^2=5.3$ for 7 degrees of
freedom (dof).

Omitting the stellar mass component gives $\chi^2=14.2$ for 8 dof;
i.e., the \chandra\ data require the stellar mass component at the
99\% level according to the F-test. If instead the DM halo is omitted,
then $\chi^2=65.7$ for 9 dof, showing that the \chandra\ data require
it at the $\approx 4\sigma$ level. This X-ray evidence for DM in \src\
is noteworthy in light of recent stellar dynamical studies that yield
conflicting results for the need for a DM halo based on near-IR data;
i.e., Y17 do not require DM while \citet{wals17a} do require it.

For reference, the best-fit virial radii are:  $\rtwofiveh=150\pm 17$~kpc,
$\rfiveh=295\pm 38$~kpc, and $\rtwoh=429\pm 56$~kpc. The extent of the
data is $\approx\rtwofiveh/2$ which we indicate in
Figure~\ref{fig.he}. Since the outer bin is large, more relevant for
the HE models is the average bin radius $\approx 50$~kpc or $\approx\rtwofiveh/3$. 

\subsubsection{Entropy}
\label{entropy}

We plot the entropy profile in Figure~\ref{fig.he} with the entropy
scaled by $S_{500}=46.1$~keV~cm$^{2}$~\citep[see eqn.\ 3
of][]{prat10a} and give the parameter constraints in
Table~\ref{tab.he}. The entropy profile shape is similar to that of
other massive elliptical
galaxies~\citep[e.g.,][]{hump08a,hump09c,hump11a,hump12a,wern13a,buot17a}. Note
the slope $\alpha$ is somewhat shallower than the $\sim r^{1.1}$
baseline model, though the difference is only weakly significant
($\approx 1.6\sigma$).

We also show in Figure~\ref{fig.he} the baseline gravity-only
model. The entropy profile of \src\ lies well above it testifying to
the presence of non-gravitational heating. Rescaling the entropy
profile by $(f_{\rm gas}/f_{b,U})^{2/3}$, where $f_{\rm gas}$ is the
cumulative gas fraction and $f_{b,U}=0.155$ is the cosmic baryon
fraction, results in much better agreement with the gravity-only
model, especially within the region covered by the \chandra\
data. This suggests that the non-gravitational heating has not
increased the gas temperature but instead has redistributed the gas
spatially. This result is very consistent with those we have obtained
previously for the massive isolated elliptical galaxies
NGC~720~\citep{hump11a}, NGC~1521~\citep{hump12b}, and NGC~6482 (B17)
and results for galaxy clusters~\citep[e.g.,][]{prat10a}.

We mention that we investigated adding a break radius to the entropy (see equation 3 of B17) 
and found the data did not require it. Including such a break yielded an
entropy profile and overall HE solution very consistent with the
no-break case. 

\subsubsection{DM Profile}
\label{dm}

While the DM halo is clearly required (\S \ref{overview}), the data
do not distinguish between profiles with central cusps (NFW, Einasto)
and cores (logarithmic potential $\ln(r_c^2 + r^2)$).  Moreover, fits
using the NFW and Einasto profiles are indistinguishable and yield
similar parameter values. These results are fully consistent with what
we found for NGC~6482 (B17). 

\subsubsection{SMBH}
\label{smbh}

While we fixed the SMBH mass by default in our models, we found that
the \chandra\ data were able to constrain $M_{\rm BH}$, albeit
weakly. The key reason why only weak constraints are possible with the
present data is that the central aperture has a
radius $1\farcs 7$ whereas the SMBH radius of influence is $r_g\approx
0\farcs 5$. Using a flat prior for $M_{\rm BH}$ over the range
$(0.3-20)\times 10^9\, M_{\odot}$ we obtain $M_{\rm BH} = (5\pm 4)\times
10^{9}\, M_{\odot}$, with a  90\% upper limit of
$M_{\rm BH} = 1.4\times 10^{10}\, M_{\odot}$, very consistent with the
recent stellar dynamical measurement by \citet{wals17a}.

If instead we use a flat prior on the logarithm, the SMBH is not
detected and a more stringent upper limit is indicated:
$M_{\rm BH} = (1.4\pm 1.7)\times 10^{9}\, M_{\odot}$, with a 99\%
upper limit of $M_{\rm BH} = 9.4\times 10^{9}\, M_{\odot}$. The
sensitivity of $M_{\rm BH}$ to the prior shows that the parameter is
not constrained robustly by our bayesian analysis. A standard
frequentist $\chi^2$ fit yields
$M_{\rm BH} = (3.9\pm 2.6)\times 10^{9}\, M_{\odot}$ more in line with
the result for the flat (non-logarithm) prior. We conclude that the
present data are consistent with the $M_{\rm BH}$ determination by
\citet{wals17a}, and improvement in the constraint awaits
precise measurements of the hot ISM properties in a smaller aperture
closer to $r_g$. 

We note that omitting the SMBH from the HE models has a negligible
impact on the quality of the fit. Consequently, the centrally peaked
temperature profile (Figure~\ref{fig.best}) is not caused by the
gravitational influence of the SMBH, nor is it due to hard emission
from an AGN (\S \ref{spec}).  The centrally peaked temperature
profiles observed in several massive elliptical galaxies (e.g.,
NGC~6482, B17) may be explained by classical wind
models~\citep[e.g.,][]{davi91a,ciot91a}.

\subsubsection{Stellar Mass and IMF}
\label{stars}

The result we obtain for the stellar mass-to-light ratio,
$\mlhband = 1.33\pm 0.21$ solar, agrees very well with the
stellar-dynamical analysis of \citet{wals17a} who found
$\mlhband = 1.3\pm 0.4$ solar. \citet{wals17a} report their value also
agrees with that of Y17 and is consistent with single-burst stellar
population synthesis models with either a Kroupa ($\mlhband = 1.2$
solar) or Salpeter ($\mlhband = 1.7$ solar) IMF. However, the smaller error bar
we obtain for $\mlhband$ remains fully consistent with the Kroupa IMF
but is marginally inconsistent $(\approx 2\sigma)$ with the Salpeter
IMF. 

The agreement with a Kroupa (or Chabrier) IMF we find for \src\ is
typical for X-ray HE studies of massive elliptical galaxies (see
discussion in \S 8.3 of B17). We also note that our result for \src\
is not dependent on our use of the accurate MGE model of the $H$-band
light (\S \ref{he}). If instead we use a de Vaucouleurs model with the
half-light radius from the Two Micron All-Sky Survey (2MASS) Extended
Source Catalog~\citep{jarr00a}, we obtain $\mlhband = 1.24\pm 0.19$
solar, very consistent with the MGE result.

Finally, we note that the good agreement of the values of $\mlhband$
obtained by us and \citet{wals17a} supports the accuracy of the
mass-measurement techniques used by both studies; i.e., in our case,
the accuracy of the hydrostatic equilibrium approximation for
\src. Note also that the consistency of the stellar mass supports the
stellar mass-size relation for CEGs obtained by Y17 indicating that
the structure of the CEGs matches the redshift $\sim 2$ red nugget
population rather than the low-redshift ETG population.

\subsubsection{Density Slope and DM Fraction}
\label{slope}


\begin{table}[t] \footnotesize
\begin{center}
\caption{Mass-Weighted Total Density Slope and DM Fraction}
\label{tab.slope}
\begin{tabular}{rrcc}   \hline\hline\\[-7pt]
Radius & Radius\\
(kpc) & ($R_e$) & $\langle\gamma\rangle$ & $f_{\rm DM}$ \\
\hline \\[-7pt]
  2.3 &   1.0 & $ 2.22 \pm  0.08$ & $ 0.20 \pm  0.07$ \\
  4.6 &   2.0 & $ 2.07 \pm  0.09$ & $ 0.38 \pm  0.09$ \\
 9.2 &   4.0 & $ 1.90 \pm  0.07$ & $ 0.60 \pm  0.07$ \\
 23.0 &  10.0 & $ 1.87 \pm  0.13$ & $ 0.82 \pm  0.03$ \\
\hline \\
\end{tabular}
\tablecomments{The mass-weighted slope is evaluated for the fiducial
  HE model using equation (2) of~\citet{dutt14b}. The DM fraction is
  defined at each radius $r$ as, $f_{\rm DM} = M_{\rm DM}(<r)/M_{\rm total}(<r)$.} 
\end{center}
\end{table}

%

Y17 report an average total mass density slope
$\langle\gamma\rangle = 2.3$ within $r=R_e$ for their sample of 16
CEGs. This modestly exceeds the average slopes of normal massive local ETGs
($2.15 \pm 0.03$, intrinsic scatter 0.10) obtained by \citet{capp15a}
also within $R_e$. In Table~\ref{tab.slope} we list mass-weighted
slopes evaluated for several radii. We obtain $\langle\gamma\rangle =
2.22\pm 0.08$ within $r=R_e$, which agrees very well with the average
CEG value from Y17 and is also consistent with the local ETGs. 

Both the average slope we obtain for \src\ within $R_e$ and its
variation with radius broadly agree with the average results of Y17
for CEGs. The CEGs reported
by Y17 have average slopes that decrease with radius from a value of 2.3 within
$1R_e$ to 1.99 at larger radius. As seen in
Table~\ref{tab.slope}, $\langle\gamma\rangle$ for \src\ decreases with
increasing radius out to $10R_e$ and (not shown) begins to increase
soon after. For comparison, the instantaneous slope (i.e., not
mass-weighted) is $\approx 2.3$ near the center, reaches a minimum
value of $\approx 1.7$ at $r\approx 3R_e$ and increases for larger
radii approaching the slope of 3 for the NFW profile. \citet{capp15a}
find an average slope of $2.19 \pm 0.03$ with 0.11 scatter over
$0.1R_e-4R_e$ for normal ETGs which is significantly larger than the
mass-weighted slope we measure within $4R_e$ ($1.90 \pm 0.07$). Hence,
like the CEGs studied by Y17,
we obtain for \src\ smaller density slopes than the normal ETGs for
radii larger than $R_e$ out to $\approx 4R_e$. (We note that the slope
of NGC~6482 (B17) is consistent with the normal ETGs.)

Turning to the DM fraction, Y17 obtain $f_{\rm DM} = 0.11$ within
$r=R_e$ for the CEGs which is lower than the value 0.19 of the normal
ETGs studied by \citet{capp15a} and accounting for the higher mass range of
the CEGs (see Y17). We obtain $f_{\rm DM} =  0.20 \pm  0.07$
(Table~\ref{tab.slope}) for \src\ which agrees very well with the
normal ETGs and is nearly double the value of the CEGs, though the
difference is only weakly significant. 


Finally, the slope-$R_e$ relation (\citealt{hump10a}; see also
\citealt{auge10a}),
$$\gamma=2.31 - 0.54\log (R_e/\rm kpc)$$
predicts an average slope 2.11 for \src\ over 0.2-10$~R_e$. The
mass-weighted slope we obtain (Table~\ref{tab.slope}) is
$\approx 11\%$ below the predicted value but within the observed
scatter~\citep{auge10a}; note B17 found NGC~6482 had a slope
$\approx 12\%$ above the predicted value.

\subsubsection{Halo Concentration and Mass}

Whereas Y17 and \citet{wals17a} were unable to constrain the DM halo
concentration using stellar dynamics, we obtain interesting
constraints, $c_{200}=17.5\pm 6.7$ and
$M_{200} = (9.6\pm 3.7)\times 10^{12}\, M_{\odot}$
(Table~\ref{tab.he}), despite the short \chandra\ exposure.  These
best-fitting values exceed the value $c_{200}=6.6$ of the mean
$c_{200}-M_{200}$ relation from \lcdm\ by
$\approx 4\sigma$~\citep{dutt14a}. While the discrepancy is only
$\sim 2\sigma$ significant in terms of the measurement error, the
large $c_{200}$ may provide evidence for weak adiabatic contraction as
we argued for NGC~6482; i.e., the ``forced quenching'' model of
\citet{dutt15a} implemented as the AC4 model in B17 yields a similar
$M_{200}$ and a smaller $c_{200}=15$ that is less discrepant
($\approx 3\sigma$) with the mean \lcdm\ relation (and does not alter
the best-fitting $\mlhband$.)

In fact, the concentration discrepancy may be even larger for
\src. Unlike the results quoted previously for other model parameters
(e.g., $\mlhband$) , we find that the concentration values differ by
$>1\sigma$ error depending on how we define the Bayesian best-fitting
value.  Above we have focused on the ``Best Fit'' values~(see \S
\ref{overview}).  For well-constrained parameters, the ``Best Fit''
and ``Max Like'' parameter values closely correspond; e.g., see the
results for NGC~6482 (B17) and RXJ~1159+5531~\citep{buot16a}.  But the
concentration and virial mass, which are global halo parameters, are
not very well constrained for \src\ since the \chandra\ measurements
of gas temperature and density currently extend only out to an average
binned radius $\approx \rtwofiveh/3$.

The Max Like values we obtain are, $c_{200}=25.9$ and
$M_{200} = 5.1\times 10^{12}\, M_{\odot}$, where $c_{200}$ is over
$5\sigma$ above the mean \lcdm\ relation, almost as discrepant as
NGC~6482 (B17). We note also that the Max Like parameters closely
correspond to those obtained from a standard frequentist $\chi^2$ fit;
i.e., $c_{200}=25.5$ and $M_{200} = 5.4\times 10^{12}\, M_{\odot}$.

\subsubsection{Gas and Baryon Fraction}
\label{baryfrac}

Results qualitatively similar to the concentration are obtained for
the global baryon fraction $(f_{\rm b,200})$. While the mean Best Fit
value $(f_{\rm b,200} = 0.070\pm 0.045)$ is less than half (and
$\approx 2\sigma$ below) the cosmic value (0.155), the Max Like value
(0.14) is fully consistent with it. Which of these two values of the
baryon fraction better approximates reality is important when
considering the ``Missing Baryons Problem'' at low
redshift~\citep{fuku98}. If the higher value prevails it would lend
support to the notion that, at least in massive elliptical galaxy /
small group halos, most of the baryons could be located in the outer
halo as part of the hot component -- consistent with our results
obtained previously for NGC~720, NGC~1521, and
NGC~6482~\citep{hump11a,hump12b,buot17a};

Note that we do not estimate the contribution to $f_{\rm b,200}$ from
smaller, non-central galaxies as we have done in previous studies
since their contribution is expected to be negligible compared to the
large statistical errors obtained already for $f_{\rm b,200}$.

\subsection{Error Budget}
\label{sys}

We have considered an extensive number of systematic tests and
examined their impact on the measured parameters of the HE model. Some
of these have been discussed in previous sections; e.g., adding a
break radius to the entropy profile (\S \ref{entropy}), using a de
Vaucouleurs profile of the stellar light (\S \ref{stars}). We
considered most of the systematic tests discussed in \S 7 of
B17. However, due to the relatively large statistical errors on the HE
parameters for \src, we find all of those systematic errors to be
negligible in that they are less than the $1\sigma$ statistical errors.
Consequently, in this section we only summarize a few notable tests
associated with choices in the spectral analysis (\S \ref{specmod} and
\S \ref{spec}) and the treatment of the plasma emissivity in the HE
models (\S \ref{he}). All of these tests had no significant effect on
the HE parameters. 

\medskip

\noindent{\it Constant Metallicity: } Whereas the results we have presented
allow the metallicity to vary with radius, we also considered the
constant metallicity solution reported in \S \ref{spec}. 

\smallskip 

\noindent{\it Soft CXB: } We examined adjusting the fluxes of the soft
CXB components by factors of 0.5 and 2.

\smallskip 

\noindent{\it Unresolved LMXBs: } We examined adjusting the nominal flux of
the unresolved LMXB component by factors of 0.5 and 2.

\smallskip 

\noindent{\it Plasma Emissivity: } The plasma emissivity
$\Lambda_{\nu}(T,Z)$ in our fiducial HE model is evaluated
self-consistently at any radius using the temperature of the
model. The metallicity used to evaluate $\Lambda_{\nu}(T,Z)$ at any
radius is obtained by fitting a projected, emission-weighted smooth
model to the measured metallicity profile (Table~\ref{tab.gas}). The
smooth model we use is essentially a $\beta$ model plus a
constant. For comparison, we also adopted the procedure we have
favored in our previous studies (e.g., see \S 4 of B17) of
interpolating the radial grid established by the measured binned
metallicities in projection as a proxy for the three-dimensional
metallicity profile.

\section{Discussion and Conclusions}
\label{conc}

We have found for \src\ that the entropy profile and global mass
properties ($c_{200}, M_{200}$) are very similar to those of the
fossil group NGC~6482 (B17) and the massive, isolated nearly
fossil\footnote{Although NGC~720 and NGC~1521 are typically classified
  as members of larger groups owing to more distant galaxy
  associations, we refer to them as ``nearly'' fossil systems since
  they obey the fossil classification within their projected virial
  radii.}  systems NGC~720~\citep{hump11a} and
NGC~1521~\citep{hump12b}.  Although for \srctwo\ the \chandra\ data
did not allow for detailed HE analysis, we can place its X-ray
properties in context, along with those of \src, through comparison with
global X-ray scaling relations.

The ETG scaling relations for local galaxies seriously underpredict
the X-ray luminosities we have measured for \src\ and \srctwo. The
$L_{\rm x}-T_{\rm x}$ relation reported by \citet{goul16a} for the
most massive nearby ellipticals predicts
$L_{\rm x}\approx 7\times 10^{40}$~erg~s$^{-1}$ in the 0.3-5.0~keV
band for \src\ while we measure a value
$\sim 1\times 10^{42}$~erg~s$^{-1}$ in the same band that is a factor
of $\sim 14$ times larger. For \srctwo\ the difference is a factor of
$\sim 3$. The $L_{\rm x}, L_{\rm K}, \sigma_e$ scaling relation of
\citet{goul16a} yields even larger factors\footnote{In these
  comparisons we have used the ``full aperture'' scaling relations of
  \citet{goul16a} defined within the radius where the diffuse hot gas
  flux equals the background. For \src\ this corresponds to Annulus~7
  and leads to an $L_{\rm x}\approx 0.95\%$ of the value at 100~kpc
  listed in Table~\ref{tab.prop}. For \srctwo\ the corresponding
  radius is $\approx 0\farcm 3$ and $L_{\rm x}\approx 0.6$ of the
  value at 100~kpc.}. The fact that \src\ and \srctwo\ have
$L_{\rm x}$ values that greatly exceed the scaling relations of normal
ETGs is unsurprising since the total masses inferred for these
galaxies $(\sim 10^{13}\, M_{\odot})$ indicate group-scale halos (even
though both galaxies are rather isolated -- e.g.,
\citealt{ferr17a}). Indeed, using the $L_{\rm x}-T_{\rm x}$ results
for galaxy groups by \citet{lovi15a} with the average temperatures we
have measured (Table~\ref{tab.prop}) yields good agreement for \src\
but overpredicts $L_{\rm x}$ for \srctwo\ by a factor of $\sim 5$;
i.e., \srctwo\ lies roughly midway between the scaling relations for
ETGs and groups.

If these CEGs are indeed largely untouched descendents from the
$z\sim 2$ population of red nuggets~(\citealt{ferr17a}; Y17) that
undergo little ``phase 2'' stellar accretion over that time, it is
remarkable that they are each the dominant central galaxy in a
group-size halo. Apparently all the merging in these groups
occurred in the assembly of the red nugget, making these systems truly
ancient fossil groups.

The results we have presented in this paper were first summarized in
observing proposals submitted to the \chandra\ AO19 and \xmm\ AO17
calls for proposals in 2017. Both proposals were approved for deep
follow-up observations of \src.

\acknowledgements 

We thank the anonymous referee and Dr.\ A.\ Y{\i}ld{\i}r{\i}m for comments on
the manuscript. DAB gratefully acknowledges partial support from the
National Aeronautics and Space Administration (NASA) under Grant
NNX15AM97G issued through the Astrophysics Data Analysis Program. This
research has made use of the NASA/IPAC Extragalactic Database (NED)
which is operated by the Jet Propulsion Laboratory, California
Institute of Technology, under contract with the National Aeronautics
and Space Administration.

\bibliographystyle{apj}

\end{document}